# Magnetic skyrmion artificial synapse for neuromorphic computing

Kyung Mee Song,[1] Jae-Seung Jeong,[2] Biao Pan,[3] Xichao Zhang,[4] Jing Xia,[4] Sun Kyung Cha,[1] Tae-Eon Park,[1] Kwangsu Kim,[1,5] Simone Finizio,[6] Jörg Raabe,[6] Joonyeon Chang,[1,7] Yan Zhou,[4] Weisheng Zhao,[3] Wang Kang,[3] Hyunsu Ju,[2] Seonghoon Woo[8*]

[1]*Center for Spintronics, Korea Institute of Science and Technology, Seoul 02792, Korea*
[2]*Center for Opto-Electronic Materials and Devices, Korea Institute of Science and Technology, Seoul 02792, Korea*
[3]*Fert Beijing Institute, BDBC, and School of Microelectronics, Beihang University, Beijing 100191, China*
[4]*School of Science and Engineering, The Chinese University of Hong Kong, Shenzhen, Guangdong 518172, China*
[5]*Department of Physics, University of Ulsan, Ulsan 44610, Korea*
[6]*Swiss Light Source, Paul Scherrer Institut, 5232 Villigen, Switzerland*
[7]*Department of Materials Science & Engineering, Yonsei University, Seoul 03722, Korea*
[8]*IBM T.J. Watson Research Center, 1101 Kitchawan Rd, Yorktown Heights, New York 10598, USA*

*Correspondence: shwoo@ibm.com*

Since the experimental discovery of magnetic skyrmions achieved one decade ago[1], there have been significant efforts to bring the virtual particles into all-electrical fully functional devices, inspired by their fascinating physical and topological properties suitable for future low-power electronics[2]. Here, we experimentally demonstrate such a device – electrically-operating skyrmion-based artificial synaptic device designed for neuromorphic computing. We present that controlled current-induced creation, motion, detection and deletion of skyrmions in ferrimagnetic multilayers can be harnessed in a single device at room temperature to imitate the behaviors of biological synapses. Using simulations, we demonstrate that such skyrmion-based synapses could be used to perform neuromorphic pattern-recognition computing using handwritten recognition data set, reaching to the accuracy of ~89%, comparable to the software-based training accuracy of ~94%. Chip-level simulation then highlights the potential of skyrmion synapse compared to existing technologies. Our findings experimentally illustrate the basic concepts of skyrmion-based fully functional electronic devices while providing a new building block in the emerging field of spintronics-based bio-inspired computing.

## I. INTRODUCTION

Magnetic skyrmions are topologically nontrivial swirling spin textures that exhibit fascinating physical characteristics and have considerable potential as the basis for highly energy-efficient data storage, processing and transmission devices[2]. Skyrmions were first observed at ultralow temperatures in magnetic compounds such as MnSi[1] and FeCoSi[3], whose non-centrosymmetric B20 crystal structure gives rise to an anti-symmetric exchange interaction between neighboring spins, called Dzyaloshinskii-Moriya interaction (DMI)[4,5]. It has also been found that skyrmions can exist at room temperature in sputtered thin films and multilayers such as Ta/CoFeB/TaOx[6] and Pt/CoFeB/MgO[7], stabilized by interface-oriented DMI, perpendicular magnetic anisotropy (PMA), and stray fields. More recently, their electrical creation[8–11], motion[6–8,12], detection[13,14] and deletion[11] have been separately demonstrated in various material platforms, suggesting that it might be possible to create an all-electrical fully functional skyrmion-based electronic device that has remained elusive so far. Here, we present such electronic device for the first time, where skyrmions are written, driven, read and erased electrically on a single device at room temperature.

In particular, we demonstrate a skyrmion-based artificial synaptic device designed for neuromorphic computing and thus artificial neural network (ANN), which could be used in broader technology fields in the future. Neuromorphic computing, inspired by the human brain's biological nervous system, has recently attracted significant attention across diverse scientific and technology areas. Extremely low-power consumption is achievable in such system due to the massively paralleled nature operated by neurons (computing elements) and synapses (memory elements)[15]. Since the analog memory capability is required for synaptic operations, several existing non-volatile memory (NVM) technology have attempted to emulate the biological synapse including phase-change[16] and resistive[17] devices. Spintronics-based multi-leveled synaptic devices have also been suggested using magnetic domain walls[18], where the displacement of domain walls could generate multi-resistance states across magnetic tunnel junctions (MTJs). However, the relatively large threshold current for domain wall motion and their stochastic pinning/depinning nature may limit the device performances. Unlikely, skyrmions show rigid-body and particle-like behaviors[19], so that multiple nanoscale skyrmions can accumulate within a defined device area without interacting with topographic defects, exhibiting potential for analog storage devices as recently simulated in ref. [20]. In this work, we experimentally demonstrate that skyrmions can be used for neuromorphic computing by realizing such skyrmion-based artificial synapse. Together with the recent demonstrations of new computing paradigms based on spintronic devices, including magnetic ANNs using spin-torque nano oscillators[21,22] and non-conventional probabilistic computing architecture using thermally induced skyrmion dynamics[23,24], we believe this work also constitutes a major advance in spintronics-based bio-inspired computing.

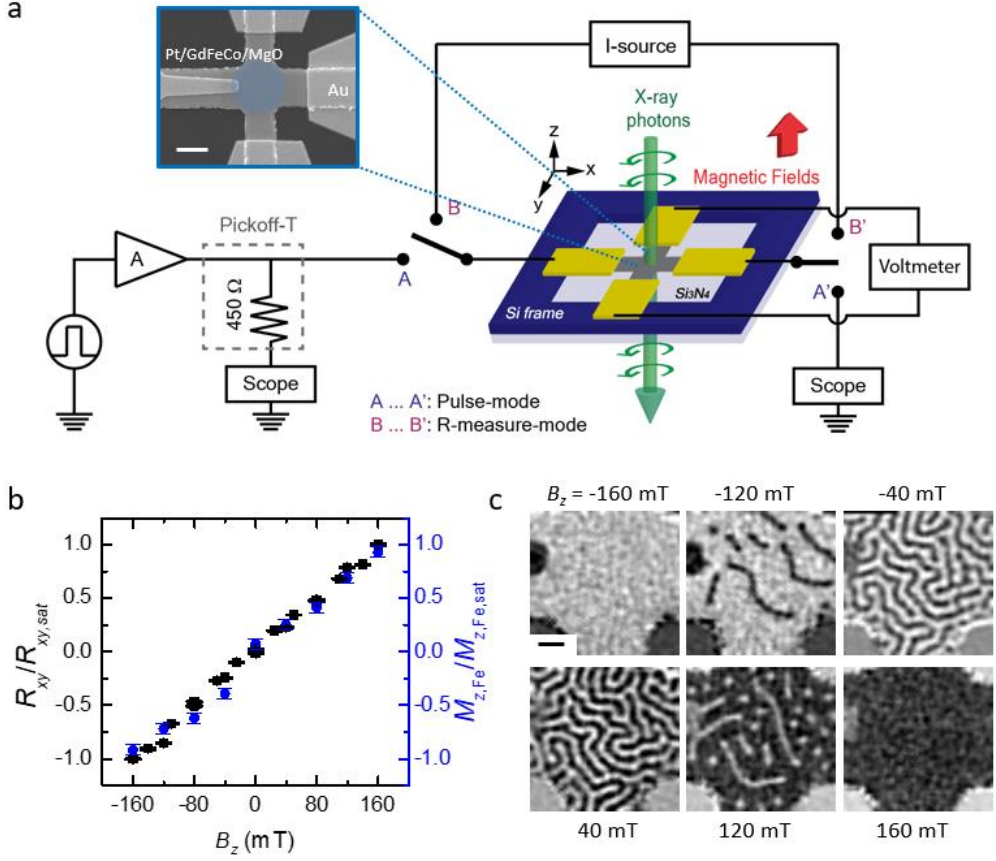

**Figure 1. Schematic of the experimental set-up and X-ray imaging of domain structures. a,** Schematic of scanning transmission X-ray microscopy (STXM) and simultaneous electrical measurement geometry. Scanning electron microscope image of the measured device is enclosed, and the marked circular area denotes the effective area where the Hall bar resistance was measured [see Methods and Supplementary Fig. S2 for details]. The A-A' connection is used for electrical pulse applications along the horizontal stripe (along *x*-axis) of the Hall bar geometry, while the B-B' connection is used for subsequent resistance measurements along the vertical (along *y*-axis) Hall bar contacts. Scale bar, 4 µm. **b,** The measured Hall resistance ($R_{xy}/R_{xy,\text{sat}}$) and the extracted out-of-plane magnetization of Fe atoms ($M_{z,\text{Fe}}/M_{z,\text{Fe,sat}}$) as a function of applied magnetic field, $B_z$. Error bars indicate the standard deviation of 10 individual resistivity measurements ($R_{xy}/R_{xy,\text{sat}}$) and spatial resolution and counting statistics ($M_{z,\text{Fe}}/M_{z,\text{Fe,sat}}$). **c,** Exemplary STXM images acquired at several out-of-plane magnetic fields, $B_z$. Dark and bright contrast correspond to Fe-magnetization oriented up (along +*z*) and down (along -*z*), respectively. Note that Fe (Gd) moments are aligned (anti-)parallel to magnetic fields as the measurements were conducted above the magnetization compensation temperature, $T_M \approx 190$ K, of this material as discussed in Supplementary Fig. S3. Scale bar, 1 µm.

## II. RESULTS

### A. STXM imaging on ferrimagnetic heterostructures

[Pt (3 nm)/Gd$_{24}$Fe$_{66.6}$Co$_{9.4}$ (9 nm)/MgO (1 nm)]$_{20}$ ferrimagnetic multilayer stacks (Pt/GdFeCo/MgO hereafter) with PMA were studied with high-resolution scanning transmission X-ray microscopy (STXM) [see Methods and Supplementary Fig. S1 for details]. This asymmetric ferrimagnetic multilayer stack is known to have reasonably large DMI, $|D| = 0.98$ mJ m$^{-2}$, where ~200 nm-size skyrmions can be stabilized at room temperature[12]. Moreover, electrical writing/deleting from natural sites[11] and efficient skyrmion motion with a reduced skyrmion Hall effect[12] have been individually observed in this ferrimagnetic stack in our previous studies, which are combined together to realize fully functional skyrmion devices in the present work.

Figure 1a first shows the experimental set-up used for the simultaneous resistance measurement and domain imaging. Two circuits for electrical pulse application and the device resistance measurement were alternately used. Magnetic domain images of the device were also recorded after each pulse application by probing the transmitted X-ray intensity at the Fe $L_3$ absorption edge, where X-ray magnetic circular dichroism (XMCD) provides contrasts corresponds to the out-of-plane projection of the magnetization. A scanning electron microscopy (SEM) micrograph of the measured device is included in Fig. 1a, where the blue-colored circular area denotes the effective synaptic device region where the resistance variations upon magnetization reversal are measured in this Hall-cross geometry [see Supplementary Fig. S2 for the details on active device area calculation].

Figure 1b and 1c show the measured Hall resistances and magnetic domain configurations as a function of out-of-plane magnetic field, $B_z$, respectively. The same analysis method used in [ref. [14]] was employed for the investigation of the skyrmion-induced change in the Hall resistivity. In Fig. 1b, $R_{xy}$ refers to the measured Hall resistance and $R_{xy,\text{sat}}$ refers to the Hall resistance at the saturated state, which corresponds to a Hall bar with a completely black XMCD contrast at $B_z = 160$ mT. Figure 1b also plots the normalized out-of-plane magnetization of Fe atoms, $M_{z,\text{Fe}}/M_{z,\text{Fe,sat}}$, computed from the domain area of STXM images at each magnetic field (Fig. 1c) [see [14] and Methods for details]. It should be noted that the magnetic moment of Fe (Gd) aligns (anti-)parallel with the external magnetic field at room temperature, which is higher than the measured magnetization compensation point of this material, $T_M \approx 190$ K [see Supplementary Fig. S3 for details]. Both normalized Hall resistance ($R_{xy}/R_{xy,\text{sat}}$) and out-of-plane magnetization ($M_{z,\text{Fe}}/M_{z,\text{Fe,sat}}$) exhibit a linear dependence with respect to the applied magnetic field, indicating that the measured resistances are reflected by the changes in the out-of-plane magnetic configuration.

The Hall resistivity of such magnetic material is given by the sum of the ordinary ($\rho_{xy}^N$), the anomalous ($\rho_{xy}^A$) and the topological ($\rho_{xy}^T$) Hall resistivities[13,14,19], the magnitude of topological Hall effect (THE) could be suppressed in ferrimagnetic multilayers because the THE contributions from the two sublattices are destructive due to their antiferromagnetically coupling[25]. Even considering the presence of a THE contribution, as magneto-transport properties may be dominated by the participating electrons of FeCo because the $4f$ shell of Gd locates below the Fermi energy level, this will still be giving a contribution to the total Hall resistivity that depends on the number of skyrmions present in the active area of the sample. This provides flexibility to our design, allowing its use independently on whether the skyrmions are stabilized in a ferri- or ferromagnetic material. We experimentally measured the ordinary Hall resistance contribution and confirmed its negligible contribution [see Methods for details]. With these established electrical measurement and X-ray magnetic imaging technique, we next examine the behaviors of this artificial synapse.

**B. Electrical operation of magnetic skyrmion artificial synapse**

Figure 2a illustrates the schematic of a skyrmion-based artificial synapse and its working principle, and Néel-type skyrmions mapped onto spheres are schematically shown in three-dimensional space. In a skyrmion synapse, the synaptic weights are proportional to the number of skyrmions in a synapse. Therefore, the electric current-controlled accumulation and dissipation of skyrmions within an active synaptic device area can imitate the linear variations of synaptic weights during potentiation and depression, respectively. Figure 2b demonstrates the experimentally measured electrical operation of a skyrmion synapse, where the corresponding magnetic configuration of each resistance state is also imaged by STXM as shown in Fig. 2c. After applying the initial saturating magnetic field of $B_z = 160$ mT, the field was reduced to $B_z = 140$ mT (image #1 in Fig. 2c). Although this magnetic field still provides enough energy to keep the saturated magnetization, the reduced field opens a room for metastable state, where skyrmions could remain stable once they are generated due to the annihilation energy barrier[26,27].

During potentiation, each image shown in Fig. 2c was acquired after injecting current pulses with the gradually increasing current-density amplitudes of $1.69 \times 10^{10}$ A m$^{-2}$ < $|j_a|$ < $4.24 \times 10^{10}$ A m$^{-2}$ and pulse duration of 100 ns, with the polarity as indicated in Fig. 2b inset. The applied unipolar pulses during both potentiation and depression were designed and optimized to generate/delete the controlled number of skyrmions using current-induced spin-orbit torques (SOTs) as investigated in our previous study[11], assisted by current-induced joule heating that contributes to the reduction of skyrmion switching thresholds[28]. In Fig. 2b, it is first noticeable that the continuous pulse application systematically generates skyrmions and drives them into the active synapse area, leading to the accumulation of skyrmions and corresponding linear resistance decrease. The first few skyrmions are generated from local pinning sites where nucleation barrier is relatively low, while the majority of skyrmions are generated near the electrical contacts and move across Hall cross area as can be seen in Fig. 2c images #1-#8. Our experiment demonstrates a total of 24 different resistivity states during potentiation. It should be noted that the skyrmion propagation direction is along the current flow direction (against electron flow), and the observed directionality agrees with the behaviors of Néel-type skyrmions with left-handed chirality[7,12], proving the topological nature of this material. We have calculated the average resistivity decrease induced by each skyrmion addition over all acquired states during potentiation, $|\Delta\rho_{xy,\text{sk-p}}| = \sum_{i=1}^{25} \frac{\Delta\rho_{xy,i}}{N_{sk,i}} / \sum_{i=1}^{25} i = 8.1 \pm 1.7$ nΩ cm, where $i$ and $N_{sk}$ denote the state number and the number of skyrmions at each state, respectively. The accumulation of a total of 41 skyrmions [see Supplementary Fig S4 for details on skyrmion number calculation] leads to the final state $\Delta\rho_{xy,\text{f}} = -3.1 \times 10^2 \pm 4.1$ nΩ cm (note that the initial state corresponds to $\Delta\rho_{xy,\text{i}} = 0$ nΩ cm).

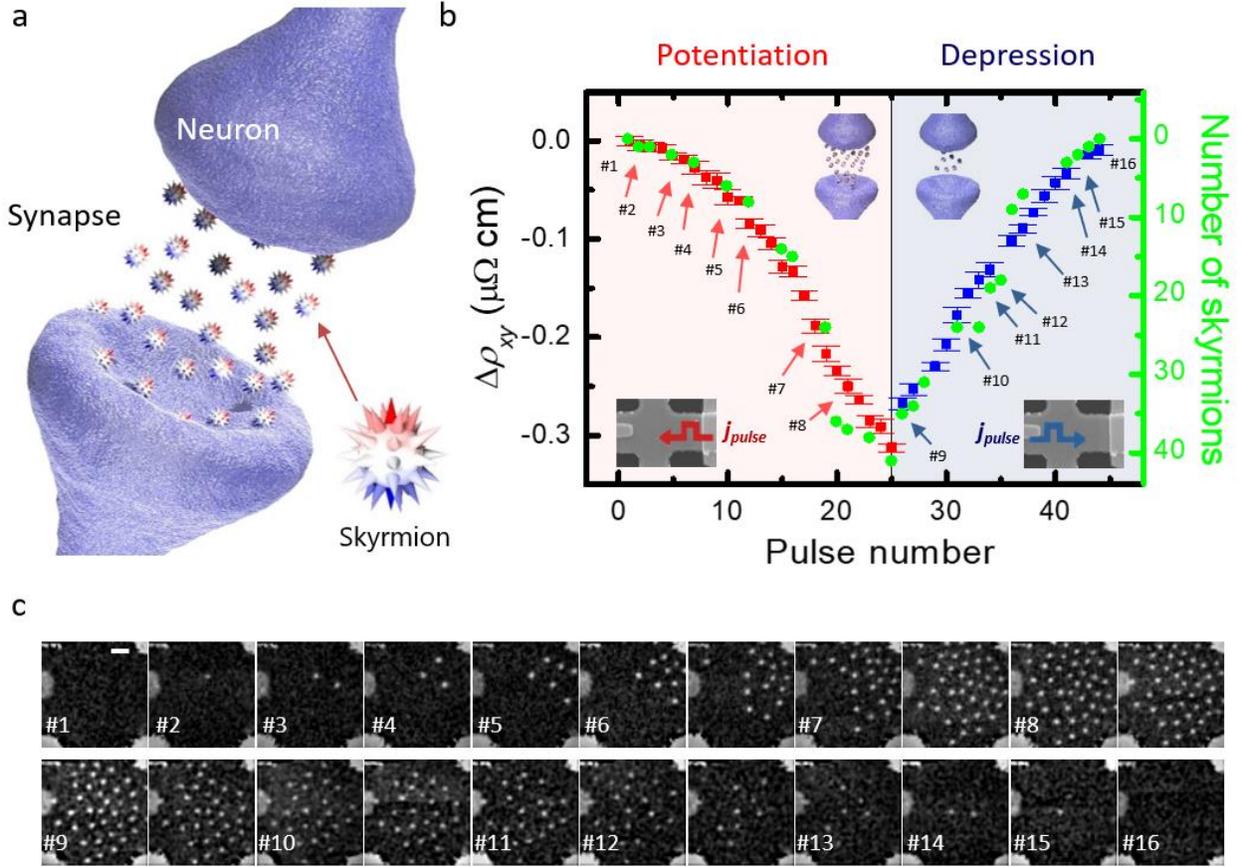

**Figure 2. Magnetic skyrmion-based artificial synapse. a,** Schematic drawing of skyrmion-based artificial synapse. In the schematic illustration, two-dimensional Néel-type skyrmions in thin films in the experiments are mapped onto spheres and are shown in three-dimensional space. The red and blue colored arrows represent magnetic moments pointing +$z$ and -$z$ directions within skyrmions, respectively. In such skyrmion synapse, the synaptic weights are proportional to the number of skyrmions, which is modulated by the electric current-controlled accumulation and dissipation of them. **b,** The measured Hall resistivity change and calculated skyrmion number as a function of injected pulse number. Note that red and blue symbols and colored areas correspond to resistivity changes (left axis) during potentiation and depression, respectively. Green symbols are used to indicate the number of skyrmions corresponding to right axis. Enclosed electrical pulses indicate the direction of charge current pulse, opposite to the direction of electron flow. Error bars denote the standard deviation of the resistivity measurements at each state. **c,** Sequential STXM images showing skyrmion populations after injecting unipolar current pulses along the track, with polarities as indicated in **b**. Upper and lower panels show domain images during potentiation and depression, respectively, and each image number, #1-#16, corresponds to each resistivity state indicated in b. Scale bar, 1 μm.

During depression, we simply reversed the polarity of gradually increasing pulses with the same amplitudes and durations (as indicated in Fig. 2b) at a slightly increased background magnetic field $B_z = 144$ mT, so that reversed pulses could excite skyrmion texture, eventually leading to skyrmion annihilation medicated by topological defects without creating additional skyrmions [11]. Note that strong inter-skyrmion repulsive forces at highly skyrmion-populated states may also contribute to the annihilation upon SOT application, as local impurities in real materials may block their synchronous motion. We indeed observe that the application of serial reversed pulses systematically annihilates skyrmions (images #9-#16) and leads to the linear resistance increase (Fig. 2b), generating 16 different states with an average resistivity change per skyrmion of $|\Delta\rho_{xy,\text{sk-d}}| = 8.9 \pm 0.1$ n$\Omega$ cm during depression, in agreement with $|\Delta\rho_{xy,\text{sk-p}}|$ measured for potentiation.

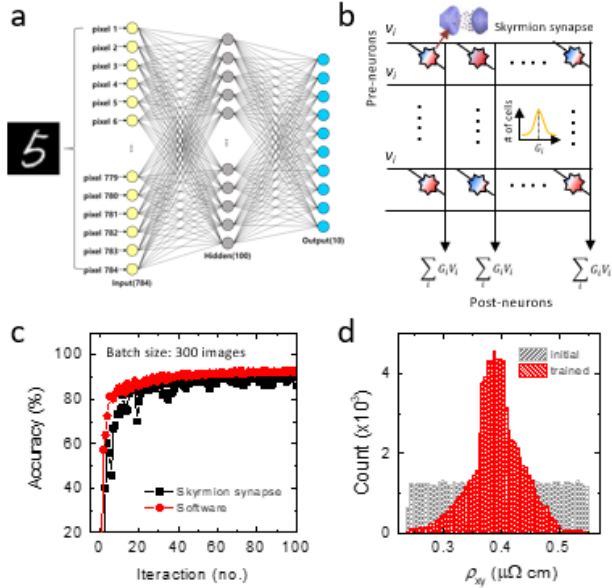

**Figure 3. The pattern-recognition simulation using skyrmion synapse. a,** The structure of tri-layer neural networks consists of 784 input neurons, 100 hidden neurons and 10 output neurons. **b,** Schematic of circuit diagram comprising skyrmion-based synapses. At each conductance level, the conductance values of skyrmion synapses follow the normal distribution with the average and the deviation, $G_i$ and $\sigma_i$, whenever synaptic weights update. **c,** Evolution of pattern recognition accuracy as a function of training iteration for an idea (software-based) and a skyrmion-based synaptic device, where the batch size of 300 images is used for each iterative training. **d,** Distribution of skyrmion synapse resistivity before- and after- trainings.

The number of skyrmions at each resistivity state is also plotted together in Fig. 2b. It is noteworthy that the measured resistivity change is slightly deviated from the corresponding skyrmion number variation especially at highly populated states, and as noted above, we believe this deviation is caused by the statistical distribution in skyrmion diameter affected by local defects and their mutual interactions. For example, at largely accumulated states (near image #8 in Fig. 2c), stronger mutual repulsive interactions between skyrmions lead to the reduction in the average skyrmion diameter, $d_{\text{sk}} = 256 \pm 25$ nm, compared to their initial distribution at sparse distribution (image #5 in Fig. 2c), $d_{\text{sk}} = 285 \pm 25$ nm [see Supplementary Fig. S5 for the details of skyrmion diameter calculation]. Considering that the skyrmion diameter is a fixed value at a defined magnetic field in a defect-free material reflecting the equilibrium at given energetic contributions [26], further material optimization could overcome the limitation and thus present complete linearity depending on the number of skyrmions. Nevertheless, the observed linear resistance variation clearly presents the key behavior of skyrmion-based artificial synapse: the current-induced analog synaptic weight modulation. We also demonstrate 'set' and 'reset' operations that are required for the programmable flexibility of an artificial synapse [see Supplementary Fig. S6 for details].

### C. Neuromorphic pattern-recognition simulation using skyrmion synapse

With these measured characteristics of skyrmion-based artificial synapse, we simulated an ANN to perform the pattern recognition, where a multilayer perceptron algorithm is used to learn the Modified National Institute of Standards and Technology (MNIST) handwritten pattern data set[29]. Figure 3 presents a simulation platform and the result of pattern recognition accuracy. Figure 3a first depicts the neural network used for learning the MNIST data consisting of 3 layers: input neural layer of 784 neurons, hidden layer of 100 neurons and output layer of 10 neurons, and skyrmion synapses are used during trainings as schematically described in Fig. 3b [see Methods for neural network simulation details]. Figure 3c shows the simulated pattern recognition accuracy as a function of training iteration for skyrmion synapse-based ANN, and the result for software-based ANN is also plotted for comparison. Our simulation demonstrates that the ANN composed of skyrmion synapses can reach ~89% pattern recognition accuracy, which is comparable to the accuracy of ideal software-based training, ~94% (Fig. 3c). Figure 3d presents the resistivity distributions translated from the synaptic weights before and after training operations. It is noteworthy that the non-ideal characteristics of demonstrated skyrmion synapses, e.g. small on/off ratio, limited number of resistance levels and cycle-to-cycle variations, may have impacted the accuracy of ANN and caused stochastic behavior observed in Fig. 3c. Although the performance of skyrmion synapse-based ANN at this stage is limited compared to other well-established technologies, we expect substantial improvement with further material and device architecture engineering. For example, the overall device characteristics of skyrmion synapses could further be optimized and enhanced, by adopting recent findings such as ~10 nm-size skyrmion at compensated ferrimagnets[30] or new electrical skyrmion-reading scheme for high ON/OFF ratio using magnetic tunnel junctions (MTJs)[31], which together would dramatically increase the scalability and electrical characteristics of skyrmion synapses.

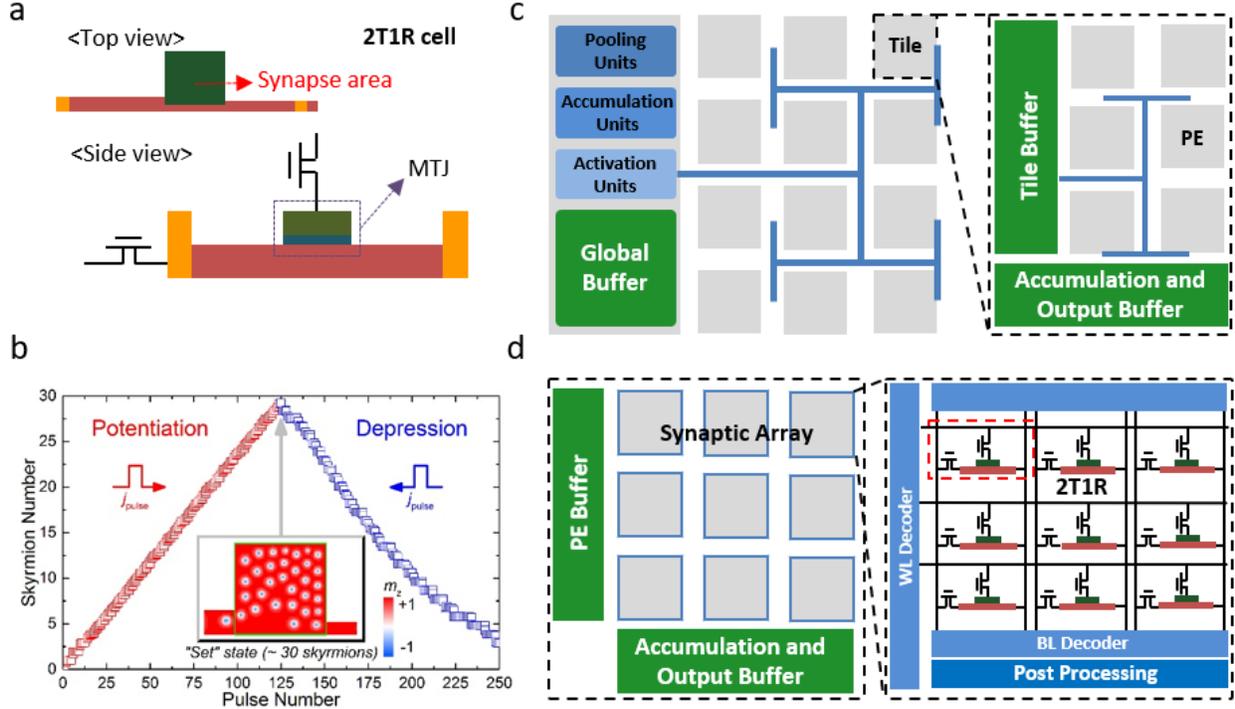

**Figure 4. The circuit implementation simulation using skyrmion synapse arrays. a,** Schematic of skyrmion synapse device in 2T1R configuration. **b,** Simulated potentiation and depression behaviors of the skyrmion synapse, where 30 skyrmions can be manipulated to generate the resistance variation in an analog fashion. Inset shows the top view of the chamber area when 125 pulses are applied. The chamber area size is 150 nm × 150 nm. The out-of-plane magnetization component is denoted by the color scale. **c,** The diagram of top level of chip architecture, which contains multiple tiles, global buffer, accumulation units, activation units (sigmoid or ReLU) and pooling units, where a tile consists of multiple processing elements (PEs). **d,** Exemplary internal architecture of a PE containing the groups of skyrmion synapse arrays, PE buffer and control units, accumulation modules and output buffer, where an example of synaptic array based on 2T1R architecture is shown[32].

## III. DISCUSSION

### A. Potential of magnetic skyrmion artificial synapse and neural network

Indeed, using additional simulations, here we discuss the potential of the skyrmion synapse-based ANN approach compared to an established technology using e.g. resistive random-access memory (RRAM). Figure 4a shows an exemplary 2T1R cell structure used for circuit implementation, where each skyrmion synapse is approximated using the known state-of-the-art chiral magnetic material parameters [see Methods for simulation parameters and other details]. In the 2T1R cell structure, one transistor is used to read resistance (synaptic weight) through the MTJ, while the other transistor is used to control electric currents along magnetic materials for skyrmion manipulation (weight update). Note that the width of the right terminal (~ 20 nm) is designed to be close to average relaxed skyrmion diameter (~7.3 nm), so that injected skyrmions from left electrode can be efficiently accumulated during potentiation. Unlike experiments that used current-induced annihilation for depression, skyrmion synapse in this case utilized the backward motion of magnetic skyrmions upon the application of reversed pulses to remove them from the active synapse area (i.e., the chamber area shown in Fig. 4b inset). Figure 4b presents the accumulation and depression behaviors of the skyrmion synapse, where a good linearity is demonstrated. It should also be noted that we could experimentally implement the MTJ with tunneling magnetoresistance (TMR) over 200 %[32,33], corresponding to $R_{on}/R_{off} \approx 3$, which is attractive for reading of multi-state skyrmion synapse. With the same pattern recognition simulation using MNIST data set performed in Fig. 3, we confirmed that such ideal skyrmion synapse-based ANN can achieve the training accuracy of ~94%, comparable to the software-based training accuracy. We then performed the circuit hardware simulation for online inference. Based on the circuit device model[20], we utilize the skyrmion synapses in a system-level approach with multi-level hierarchy. As shown in Fig. 4c, the top level of the modeled chip hierarchy consists of multiple tiles, including global buffer, accumulation units, activation units of sigmoid or ReLU (Rectified Linear Unit), and pooling units[29]. A tile contains several processing elements (PEs): tile buffer to load in neural activations, accumulation modules to add up partial sums from PEs, and output buffer. Figure 4d shows the internal structure of a PE, which is built up by the groups of skyrmion synapse sub-arrays, PE buffers, accumulation modules and output buffer. Supplementary Table S1 summarizes the evaluated chip-level performance of skyrmion synapse-based ANN and compares the performance with the RRAM-based counterpart. [see Methods for circuit simulations details], where the parameters for the RRAM device are acquired from ref. [[34]].

It is first noticeable that skyrmion synapse-based ANN in circuit level presents comparable performances compared to RRAM-based technology, and skyrmion synapse-based ANN exhibits reduced operation voltage, leading to an improved energy efficiency. For the example of weight update, RRAM (also PCM) generally requires the pulse voltage over 1 V for several nanoseconds to modulate the resistance state (synaptic weight), whereas skyrmion synapse only requires few millivolts with sub-ns pulse duration to change the skyrmion number (synaptic weight). However, the reduced on/off ratio of skyrmion synapse (<200 %) compared to RRAM case (> 1000 %) increases the energy required for dynamic reading, and further research is expected to improve the TMR of MTJ. Moreover, skyrmion synapses show relatively large bit-cell area (generally > 100 nm, as it requires multiple skyrmions to form multi-level states) and their 2T1R cell structure generates increased array area compared to RRAM synapse (10-100 nm)-based ANN. However, it is noteworthy that, within RRAM based two-terminal synapse, the weight update (training) and inference using the same voltage/current path could induce disturbance during the inference process, which, however, is not expected for skyrmion synapses with a three-terminal cell structure, where training and inference utilize separated paths.

Lastly, we discuss few other potential advantages of skyrmion synapse compared to existing RRAM or PCM-based technologies. First, skyrmion-based synapse may present good linear weight distribution (see Fig. 4b), as the weight directly depends on the number of accumulated skyrmions, while the linearity is a huge challenge for RRAM (also PCM) due to their switching mechanism[35]. Furthermore, RRAM/PCM has relatively large device-to-device and cycle-to-cycle variations owing to their intrinsic device principle as the weight depends on the temporal formation of resistance (e.g., for RRAM, the filament size and position control is challenging). However, for skyrmion synapses, the variation can be well controlled as the weight only depends on the number of skyrmions. In addition, skyrmion synapse shares many of its operation principles and material characteristics with spin-transfer torque magnetoresistive random-access memory (STT-MRAM) technology[36]. Therefore, it may offer significant endurance and retention, while RAM/PCM has a big bottleneck on endurance that limits their usage for massive training in neural network. Overall, we remark that these simulations together with our proof-of-concept experimental demonstration present a large potential of skyrmion synapse-based applications.

## IV. CONCLUSIONS

In conclusion, our results demonstrate a fully functional skyrmion-based electronic device – skyrmion-based artificial synapse – by incorporating current-induced skyrmion creation, motion, detection and deletion techniques in a single ferrimagnetic device scheme. Moreover, we present that the skyrmion synapse, where synaptic weights are modulated by the number of skyrmions in an active device area, can be implemented for the neuromorphic computing simulation of supervised pattern recognition, reaching a high learning accuracy of ~89%. We further discuss the potential of such skyrmion synapse-based ANN, which could be highly energy efficient and thus useful for advanced computing technology in the future. We believe our first experimental demonstration of the all electrical skyrmion-based device serves as the basis of skyrmion-based electrical device scheme and holds promise for applications beyond the demonstrated multi-leveled synaptic memory, including skyrmion-based racetrack memory, logic, magnonic and radio-frequency applications proposed using simulations so far[2]. Therefore, this work paves the new avenue towards skyrmion-based electronics while highlighting the possibility to use topological magnetic textures for multi-level synapse-based neuromorphic computing devices.

## V. METHODS

**Material growth and device fabrication.** The [Pt (3 nm)/Gd$_{24}$Fe$_{66.6}$Co$_{9.4}$ (9 nm)/MgO (1 nm)]$_{20}$ films were grown by DC magnetron sputtering at room temperature under 1 mTorr Ar for Pt and GdFeCo and 4 mTorr Ar for MgO at a base pressure of ~3.0 × 10$^{-8}$ Torr. A Hall bar structure with a stripe width of 4.5 μm was fabricated on a 100 nm-thick Si$_3$N$_4$ membrane wafer using electron beam lithography and lift-off technique. Ti (5 nm)/Au (100 nm) was then deposited and patterned for electrical contacts, and 200 nm-thick Al was deposited on the opposite side of the $Si_3N_4$ membrane wafer to allow for in-situ cooling during the current pulse experiments. Note that a 'tapered' geometry was used for one electrode to induce the circular current flow within the active device region that may cause circular accumulation and decumulation of skyrmions as discussed in the main text and Supplementary Fig. S2 of this manuscript. Nominally identical companion films were grown on $Si/SiO_x$ substrates to measure magnetic properties using vibrating sample magnetometry (VSM), and measurements yielded an in-plane saturation field and a net saturation magnetization of $\mu_0 H_k$ = 400 mT and $M_s$ = 1.9 × $10^5$ A $m^{-1}$, respectively. Magnetization compensation temperature, $T_M$, at which two magnetic moments from Gd and FeCo cancel each other, was acquired by performing temperature-dependent VSM measurements as shown in Supplementary Fig. S3.

**STXM measurements.** All microscope images were acquired using scanning transmission X-ray microscopy (STXM) installed at the PolLux end-station (X07DA) of the Swiss Light Source, Villigen, Switzerland. A device for STXM imaging was oriented with the surface normal parallel to the circularly-polarized X-ray beam (main text Fig. 1a), to achieve out-of-plane sensitivity through the X-ray magnetic circular dichroism (XMCD) effect. The dark and bright contrasts in all STXM images correspond to an upward ($+M_z$) and downward ($-M_z$) out-of-plane magnetization direction of the Fe atoms, respectively. The energy of the circularly-polarized X-rays was tuned to Fe $L_3$ absorption edge. The normalized out-of-plane magnetization ($M_{z,Fe}/M_{z,Fe,sat}$) plotted in Fig. 1b was calculated using the method described in ref. [14], which consisted in counting and normalizing the number of black and white pixels in the effective device area, where black (upward magnetization of Fe) and white (downward magnetization of Fe) pixels were assigned with the values of +1 and -1, respectively. For the simultaneous electrical measurements conducted together with X-ray imaging, the experiment chamber was pre-pumped down to its base pressure (about $10^{-5}$ mbar) and then filled with about 4 mbar of oxygen to prevent carbon deposition during the X-ray imaging, and to provide an additional cooling for the sample. The electrical pulse before and after sample were separately verified with -20 dB pick-off tees connected to an oscilloscope at both sides. The Hall resistance of the device was measured with a Keithley 2181A nanovoltmeter, where a Keithley 6221 d.c. source was used to apply 100 µA source current. The ordinary Hall resistivity contribution was calculated using a physical property measurement system (PPMS), by measuring the Hall resistance of the sample under an out-of-plane magnetic field of up to 2 T at room temperature. The ordinary Hall resistivity is expressed as $\rho_{xy}^N = R_0 B_z$, where $R_0$ is the ordinary Hall coefficient and $B_z$ is the applied out-of-plane magnetic field, and the measurement yielded the ordinary Hall coefficient, $R \approx -1.5 \times 10^{-11}$ Ω m $T^{-1}$, corresponding to $\rho_{xy}^N \approx$ -0.21 nΩ cm during the electrical measurement shown in Fig. 2 in main text.

**MNIST Pattern recognition simulation using experimental data.** Neuromorphic pattern recognition computing using the skyrmion synapse-based artificial neural network (ANN) was simulated as follow. Experimentally measured resistance levels and error ranges are used for the average and the deviation of each conductance level, where we selected 13 and 11 states from the measured potentiation and depression state batches, respectively. The synaptic weights of the individual skyrmion synapse were distributed randomly before training (see Fig. 3d). During the ANN simulation, at every synaptic weight update, the cell-to-cell and the update-to-update variations of the skyrmions synapses are modeled to follow a normal distribution with the average conductance and the deviation at each potentiation and depression level, denoted as $G_i$ and $\sigma_i$ in Fig. 3b in main text, respectively. Based on the inner product between the input signal vector and the synapse vector by the read current, the amount of delta weight is determined, which then provides feedback into arrays to adjust the synapse weights by inducing write or erase pulses. This process is iteratively repeated to minimize the discrepancy between the predicted and the actual labels of the MNIST data. This artificial neural network was trained with a batch size of 300 images and up to 100 iterations. The accuracy of the ANN recognition of Modified National Institute of Sciences and Technology (MNIST) data set was evaluated after every iteration.

**Skyrmion synapse simulation.** The micromagnetic simulation is carried out by using the Object Oriented MicroMagnetic Framework (OOMMF)[37], where the time-dependent spin dynamics is controlled by the Landau-Lifshitz-Gilbert (LLG) equation augmented with the antidamping-like spin transfer torque, which can be generated by the spin Hall effect in the heavy metal substrate[38,39]. The LLG equation is written as

$$\frac{d\boldsymbol{M}}{dt} = -\gamma_0 \boldsymbol{M} \times \boldsymbol{H}_{\text{eff}} + \frac{\alpha}{M_S}\left(\boldsymbol{M} \times \frac{d\boldsymbol{M}}{dt}\right) + \frac{u}{M_S}(\boldsymbol{M} \times \boldsymbol{p} \times \boldsymbol{M}) \ ,$$
(S1)

where $\boldsymbol{M}$ is the magnetization, $M_S = |\boldsymbol{M}|$ is the saturation magnetization, $t$ is the time, $\gamma_0$ is the gyromagnetic ratio with absolute value, and $\alpha$ is the Gilbert damping constant. $\boldsymbol{H}_{\text{eff}}$ is the effective field, $\boldsymbol{H}_{\text{eff}} = -\mu_0^{-1}\frac{\partial E}{\partial \boldsymbol{M}}$, where $E$ is the average energy density including the Heisenberg exchange, the perpendicular magnetic anisotropy, the applied magnetic field, the demagnetization, and the Dzyaloshinskii-Moriya interaction (DMI) energy terms. $u = \left|\frac{\gamma_0 \hbar}{\mu_0 e}\right| \frac{j\theta_{\text{SH}}}{2aM_S}$ is the antidamping-like spin transfer torque coefficient, where $\hbar$ is the reduced Planck constant, $e$ is the electron charge, $j$ is the injected current density, $\theta_{\text{SH}}$ is the spin Hall angle. $a$ is the thickness of the magnetic layer and $\boldsymbol{p} = -\hat{y}$ is the spin polarization direction.

For the micromagnetic simulation of the potentiation and depression processes, a nano-square thin-film chamber area with the side length of 150 nm is considered. A terminal area with the size of 50 nm × 40 nm is connected to the left side of the chamber. The other terminal area with the size of 50 nm × 20 nm is connected to the right side of the chamber. Note that the width of the right terminal is reduced to avoid the undesired escape of skyrmions. The thickness of the magnetic metal layer is fixed at 0.4 nm. The cell size is set as 1 nm × 1 nm × 0.4 nm. The intrinsic magnetic parameters are[40–42]: magnetization saturation $M_S$ = 500 kA/m, perpendicular magnetic anisotropy $K$ = 0.9 MJ/m$^3$, exchange constant $A$ = 15 pJ/m, DMI constant $D$ = 3.5 mJ/m$^2$, damping constant $\alpha$ = 0.3. An out-of-plane magnetic field of $B_z$ = 200 mT is applied in the chamber area in order to reduce the skyrmion size. It should be noted that this external magnetic field is not strictly required in real experiments, but skyrmion size will be a bit larger in the absence of this field.

For the potentiation process, a series of square-wave current pulses are applied to create skyrmions in the left terminal area (i.e., as the nucleation pad) and then drive skyrmions into the chamber area. Note that a localized out-of-plane magnetic field is applied at the left terminal area in a square-wave fashion, which is a numerical procedure adopted to promote the nucleation of skyrmions. In real-world materials, defects and thermal effect can naturally promote the skyrmion nucleation upon the pulse application as shown in the main text. The applied pulse length and pulse space are equal to 0.05 ns and 0.10 ns, respectively. Namely, the period of pulses equals 0.15 ns. The amplitude of current pulses is $j$ = 4×10$^{10}$ A/m$^2$ (normalized by the width of the chamber area, i.e. 150 nm). The current pulses is applied for 18.75 ns, corresponding to 125 pulses. During the pulse application, 32 skyrmions are created in the nucleation pad, of which 31 skyrmions are driven into the chamber area in a linear fashion.

For the depression process, the applied pulse current parameter and profile are the same as that of the potentiation process, however, its amplitude is switched from positive to negative at $t$ = 18.75 ns. Namely, $j$ = -4×10$^{10}$ A/m$^2$ (normalized by the width of the chamber area, i.e. 150 nm). During the pulse application, 28 skyrmions can be driven out of the chamber within 18.75 ns (i.e., 125 pulses). The left 3 skyrmions in the chamber area can be removed from the chamber area after a long time of pulse application. The last 3 skyrmions are quite difficult to be removed due to the reduced skyrmion-skyrmion interaction and small driving forces.

**MNIST Pattern recognition simulation using simulated skyrmions.** Analogous to Fig. 3 in a main text, we again adopted the neural network hardware platform NeuroSim+[29] with nonlinearity property injection, as well as DNN algorithm to perform hardware-based digital recognition through the MNIST database. The first step is software simulation (or offline training). The neural network structure we used includes 1024 neurons as input, 100 neurons as a hidden layer, and 10 neurons as an output layer, which is used for recognizing 10 number digits. The simulation trains up to 1000 epochs. Each epoch selects 8000 images randomly from 60000 training images and takes 10000 images as a testing data set. The precision of synapse weight was set to 4/5bits while the activation was set to 8bits.

**Circuit implementation simulation.** According to the network structure, the subarray size was set as 32 × 32 to fulfill the requirement of the inference. To reduce the read error rate and enhance the anti-noise ability of the device, we set the TMR to be 200%, corresponding to $R_{on}/R_{off} \approx 3$. Based on the developed circuit model, the chip simulation was conducted with a feature size of 32nm using NeuroSim+. Simulation parameters for RRAM and skyrmion synapses are listed below.

| Parameter | RRAM Synapse Value | Skyrmion Synapse Value |
|---|---|---|
| ResistanceOn | 10k | 3K |
| ResistanceOff | 100k | 9K |
| Read Voltage | 0.1 | 0.05 |
| Read Pulse Width | 1ns | 1ns |
| Access Voltage | 1.4v | 1v |
| Write Voltage | 1.8v | 30mv |
| Write Pulse Width | 1ns | 0.05ns |
| Cell Height | 5F | 14F |
| Cell Width | 23F | 44F |
| Cell bit | 5 bits | 5 bits |
| Number of States | 32 | 32 |
| Cell Type | 1T1R | 2T1R |

# ACKNOWLEDGEMENTS

This work was mainly supported by KIST Institutional Program and IBM Research. S.W. acknowledges management support from Guohan Hu and Daniel Worledge. S.W. also acknowledges Ki-Young Lee for providing artworks included in Figure 2. K.M.S., S.K.C., T.-E.P. and J.C. acknowledge the support from the National Research Council of Science and Technology (NST) (Grant no. CAP-16-01-KIST) by the Korean government (MSIP). K.K. acknowledges the support from the Basic Research Laboratory Program through the National Research Foundation of Korea (NRF) funded by the MSIT(NRF-2018R1A4A1020696). J.-S.J. and H.J. acknowledge the support from the Korea National Research Foundation program (NRF-2017R1E1A1A01077484), particularly utilized to conduct the MNIST pattern learning works of this research. J.C acknowledges the support of Yonsei-KIST Convergence Research Institute. The PolLux endstation was financed by the German Bundesministerium für Bildung und Forschung under grant No. 05KS7WE1. X.Z. acknowledges the support by the Presidential Postdoctoral Fellowship of The Chinese University of Hong Kong, Shenzhen (CUHKSZ). Y.Z. acknowledges the support by the President's Fund of CUHKSZ, Longgang Key Laboratory of Applied Spintronics, National Natural Science Foundation of China (Grant Nos. 11974298 and 61961136006), and Shenzhen Fundamental Research Fund (Grant No. JCYJ20170410171958839). W.Z. and W.K. thank the National Natural Science Foundation of China (Grant No. 61627813), the International Collaboration Project B16001 and the National Key Technology Program of China 2017ZX01032101. Parts of this work were performed at the PolLux (X07DA) endstation of the Swiss Light Source, Paul Scherrer Institut, Switzerland.

**Author contributions.**

S.W. designed, planned and initiated the study. K.M.S. grew films, fabricated devices and performed initial device characterizations. S.K.C. provided device fabrication support using electron beam lithography. S.W., K.M.S., T.-E.P, K.S.K., S.F. and J.R. performed STXM experiments at Swiss Light Source in Villigen, Switzerland. J.-S.J. and H.J. performed the neuromorphic computing simulation work. X.Z. and J.X. performed simulation on the ideal skyrmion synapse devices, and B.P., W.Z. and W.K. performed circuit implementation simulation. K.M.S., H.J. and S.W. drafted the manuscript and all authors reviewed the manuscript.

**Data and materials availability**.

All data supporting the findings of this study are available from the corresponding author upon reasonable request.

**Competing Interests**.

The authors declare no competing interests.

1. Mühlbauer, S. *et al.* Skyrmion Lattice in a Chiral Magnet. *Science* **323**, 915–919 (2009).
2. Fert, A., Reyren, N. & Cros, V. Magnetic skyrmions: advances in physics and potential applications. *Nat. Rev. Mater.* **2**, 17031 (2017).
3. Yu, X. Z. *et al.* Real-space observation of a two-dimensional skyrmion crystal. *Nature* **465**, 901–904 (2010).
4. Dzyaloshinsky, I. A thermodynamic theory of "weak" ferromagnetism of antiferromagnetics. *J. Phys. Chem. Solids* **4**, 241–255 (1958).
5. Moriya, T. Anisotropic Superexchange Interaction and Weak Ferromagnetism. *Phys. Rev.* **120**, 91–98 (1960).
6. Jiang, W. *et al.* Blowing magnetic skyrmion bubbles. *Science* **349**, 283–286 (2015).
7. Woo, S. *et al.* Observation of room-temperature magnetic skyrmions and their current-driven dynamics in ultrathin metallic ferromagnets. *Nat. Mater.* **15**, 501–506 (2016).
8. Legrand, W. *et al.* Room-Temperature Current-Induced Generation and Motion of sub-100 nm Skyrmions. *Nano Lett.* **17**, 2703–2712 (2017).
9. Hrabec, A. *et al.* Current-induced skyrmion generation and dynamics in symmetric bilayers. *Nat. Commun.* **8**, 15765 (2017).
10. Büttner, F. *et al.* Field-free deterministic ultrafast creation of magnetic skyrmions by spin–orbit torques. *Nat. Nanotechnol.* **12**, 1040–1044 (2017).
11. Woo, S. *et al.* Deterministic creation and deletion of a single magnetic skyrmion observed by direct time-resolved X-ray microscopy. *Nat. Electron.* **1**, 288 (2018).
12. Woo, S. *et al.* Current-driven dynamics and inhibition of the skyrmion Hall effect of ferrimagnetic skyrmions in GdFeCo films. *Nat. Commun.* **9**, 959 (2018).
13. Maccariello, D. *et al.* Electrical detection of single magnetic skyrmions in metallic multilayers at room temperature. *Nat. Nanotechnol.* **13**, 233 (2018).
14. Zeissler, K. *et al.* Discrete Hall resistivity contribution from Néel skyrmions in multilayer nanodiscs. *Nat. Nanotechnol.* **13**, 1161 (2018).
15. Hopfield, J. J. Neural networks and physical systems with emergent collective computational abilities. *Proc. Natl. Acad. Sci.* **79**, 2554–2558 (1982).
16. Kuzum, D., Jeyasingh, R. G. D., Lee, B. & Wong, H.-S. P. Nanoelectronic Programmable Synapses Based on Phase Change Materials for Brain-Inspired Computing. *Nano Lett.* **12**, 2179–2186 (2012).
17. Prezioso, M. *et al.* Training and operation of an integrated neuromorphic network based on metal-oxide memristors. *Nature* **521**, 61–64 (2015).
18. Lequeux, S. *et al.* A magnetic synapse: multilevel spin-torque memristor with perpendicular anisotropy. *Sci. Rep.* **6**, 31510 (2016).
19. Nagaosa, N. & Tokura, Y. Topological properties and dynamics of magnetic skyrmions. *Nat. Nanotechnol.* **8**, 899–911 (2013).
20. Huang, Y., Kang, W., Zhang, X., Zhou, Y. & Zhao, W. Magnetic skyrmion-based synaptic devices. *Nanotechnology* **28**, 08LT02 (2017).
21. Torrejon, J. *et al.* Neuromorphic computing with nanoscale spintronic oscillators. *Nature* **547**, 428–431 (2017).
22. Romera, M. *et al.* Vowel recognition with four coupled spin-torque nano-oscillators. *Nature* **563**, 230 (2018).
23. Bourianoff, G., Pinna, D., Sitte, M. & Everschor-Sitte, K. Potential implementation of reservoir computing

23. models based on magnetic skyrmions. *AIP Adv.* **8**, 055602 (2018).
24. Zázvorka, J. *et al.* Thermal skyrmion diffusion used in a reshuffler device. *Nat. Nanotechnol.* **14**, 658 (2019).
25. Barker, J. & Tretiakov, O. A. Static and Dynamical Properties of Antiferromagnetic Skyrmions in the Presence of Applied Current and Temperature. *Phys. Rev. Lett.* **116**, 147203 (2016).
26. Büttner, F., Lemesh, I. & Beach, G. S. D. Theory of isolated magnetic skyrmions: From fundamentals to room temperature applications. *Sci. Rep.* **8**, 4464 (2018).
27. Bessarab, P. F. *et al.* Stability and lifetime of antiferromagnetic skyrmions. *Phys. Rev. B* **99**, 140411 (2019).
28. Finizio, S. *et al.* Deterministic field-free skyrmion nucleation at a nano-engineered injector device. *ArXiv190210435 Cond-Mat* (2019).
29. Chen, P.-Y., Peng, X. & Yu, S. NeuroSim+: An integrated device-to-algorithm framework for benchmarking synaptic devices and array architectures, *2017 IEEE Int Electron Devices Meet. IEDM* 6.1.1–4 (2017).
30. Caretta, L. *et al.* Fast current-driven domain walls and small skyrmions in a compensated ferrimagnet. *Nat. Nanotechnol.* **13**, 1154 (2018).
31. Tomasello, R. *et al.* Electrical detection of single magnetic skyrmion at room temperature. *AIP Adv.* **7**, 056022 (2017).
32. Ikeda, S. *et al.* Tunnel magnetoresistance of 604% at 300K by suppression of Ta diffusion in CoFeB⁄MgO⁄CoFeB pseudo-spin-valves annealed at high temperature. *Appl. Phys. Lett.* **93**, 082508 (2008).
33. Wang, M. *et al.* Current-induced magnetization switching in atom-thick tungsten engineered perpendicular magnetic tunnel junctions with large tunnel magnetoresistance. *Nat. Commun.* **9**, 1–7 (2018).
34. Chen, P. & Yu, S. Compact Modeling of RRAM Devices and Its Applications in 1T1R and 1S1R Array Design. *IEEE Trans. Electron Devices* **62**, 4022–4028 (2015).
35. Moon, K., Kwak, M., Park, J., Lee, D. & Hwang, H. Improved Conductance Linearity and Conductance Ratio of 1T2R Synapse Device for Neuromorphic Systems. *IEEE Electron Device Lett.* **38**, 1023–1026 (2017).
36. Yuasa, S., Hono, K., Hu, G. & Worledge, D. C. Materials for spin-transfer-torque magnetoresistive random-access memory. *MRS Bull.* **43**, 352–357 (2018).
37. M. J. Donahue and D. G. Porter, OOMMF User's Guide, Version 1.0, Interagency Report NO. NISTIR 6376, National Institute of Standards and Technology, Gaithersburg, MD (1999) http://math.nist.gov/oommf.
38. Ralph, D. C. & Stiles, M. D. Spin transfer torques. *J. Magn. Magn. Mater.* **320**, 1190–1216 (2008).
39. Ando, K. Dynamical generation of spin currents. *Semicond. Sci. Technol.* **29**, 043002 (2014).
40. J. Sampaio, V. Cros, S. Rohart, A. Thiaville, and A. Fert, Nucleation, stability and current-induced motion of isolated magnetic skyrmions in nanostructures, *Nat. Nanotechnol*. **8**, 839 (2013).
41. R. Tomasello, E. Martinez, R. Zivieri, L. Torres, M. Carpentieri, and G. Finocchio, A strategy for the design of skyrmion racetrack memories, *Sci. Rep.* **4**, 6784 (2014).
42. X. Zhang, J. Xia, Y. Zhou, D. Wang, X. Liu, W. Zhao, and M. Ezawa, Control and manipulation of a magnetic skyrmionium in nanostructures, Physical Review B **94**, 094420 (2016).